# C-2*p* spin-polarizations along with two mechanisms in extended carbon multilayers: Insight from first principles.

Samir F Matar*

Lebanese German University. Sahel-Alma. Jounieh. Lebanon.

*Formerly at University of Bordeaux, ICMCB–CNRS. Pessac. France.

*Emails*: s.matar@lgu.edu.lb and abouliess@gmail.com

## Abstract

*From density functional theory investigations helped with crystal chemistry rationale, single-atom C, embedded in layered hexagonal $CC'_n$ (n = 6, 12, 18) networks, is stable in a magnetic state with $M(C) = 2\ \mu_B$. The examined compositions, all inscribed within the P6/mmm space group are characterized as increasingly cohesive with n, figuring mono-, bi- and tri-layered honeycomb-like C' networks respectively. The spin projected total density of states shows a closely half-metallic behavior with a gap at minority spins ($\downarrow$) and metallic majority spins ($\uparrow$). Such results together with the large C-C intersite separation and the integer $2\ \mu_B$ magnetization, let propose an intra-band mechanism of magnetic moment onset on carbon 2p states. Support is provided from complementary calculations assuming $C_2C'_{12}$ structure with planar 2 C with d(C-C)= 2.46 Å resulting into a lowering of the magnetization down to $0.985\ \mu_B$ /C atom involving a ferromagnetic order arising from interband spin polarization on C where one nonbonding spin polarizes whereas the other is involved with the bonding with the other carbon. Illustration of proofs is provided with the magnetic charge density projected onto the different atoms, showing its prevalence around C, contrary to the $C'_n$ ($C'6$ layers), as well as electron localization function ELF.*

**Introduction and context**

Unpaired electrons in outer shells of atoms let identify a paramagnetic behavior with no parallel alignment of spins. This is opposed to diamagnetic atoms where all electrons are paired such as in magnesium with $Be(2s^2) \equiv \{He\}$. Carbon, as an isolated element is paramagnetic $C(2s^2, 2p^2)$ with 2 unpaired p-electrons. In fact, such a configuration is seldom encountered in chemical systems, because carbon and its neighbors combine through pairing their respective electrons to form bonds. In paramagnetic elements, a lowering of the energy occurs when the electron spins become parallel (↑,↑) and in certain cases, below a critical temperature called the Curie temperature $T_C$, an ordering of the spins occurs adding them up constructively and leading to a long-range ferromagnetic state. Regarding the highest occupied valence states of the 1st-period transition metals, three ferromagnetic metals are found: Fe $(4s^2, 3d^6)$ with M=2.2 $\mu_B$, Co $(4s^2, 3d^7)$ with M=1.7 $\mu_B$ and Ni $(4s^2, 3d^8)$ with M=0.6 $\mu_B$; all of them having unpaired d electrons and behave as paramagnets above $T_C$. A close inspection is obtained from band structure calculations considering firstly a non-spin-polarized state (NSP). The d states show a significant localization illustrated by a high density of states (DOS) at the Fermi level $E_F$: $n(E_F)$, signaling the instability of such a spin degenerate configuration [1]. Upon carrying out further spin-polarized SP calculations accounting for two spin channels (majority spins ↑, and minority spins ↓), magnetic moments develop on the atoms. The designation of 'majority' and 'minority' arises from the fact that the former is larger in electron populations because it is energy down-shifted oppositely to the minority spins DOS which are energy up-shifted. Such properties are quantified from the quantum density functional theory DFT [2,3]. In the 2nd and 3rd transition metal periods, the broadening of the 4d- and 5d bands hinders localization of the d states and hence the onset of magnetic polarization; but under certain crystal structure conditions, a narrowing of the bands may occur such as for the 3rd period iridium in a structure characterized by isolated *IrO6* octahedra within $Ca^{II}_4Ir^{IV}O_6$ and M(Ir)= 0.5 $\mu_B$ [4].

Magnetism occurs for rare-earth gadolinium metal, characterized by half-filled 4f shell, i.e. with 7 electrons which all polarize to provide a high magnetic moment of 7 $\mu_B$ for Gd which is a ferromagnet with $T_C$ close to room temperature. The magnetic polarization is of intra-(4f)band nature, i.e. without calling for the interaction with neighboring Gd atoms. The underlying effect is that 4f states, already in the atomic state, are localized in a narrow band and keep this behavior in the organized periodic solid. Oppositely, in the ferromagnetic metals, the magnetic moment is of inter-band nature, i.e. it is mediated by the electron gas (s-like), i.e. in a "d….s….d"-like bridging and the magnetic moments are not integers as given above. To some extent, uranium presents an intermediate situation where U 5f band, less narrow than rare-earth 4f one, behaves like a transition metal's d band [5].

Then in general, the elements with outer shells having unsaturated occupations as $n$d and 4f (above examples), are likely to lead to inter- and intraband spin polarization respectively, provided the conditions of sufficient band localization are met.

Recently we extended the reasoning to light p-elements of the 1$^{st}$ period (B, C, N) and discussed the localization of the p states due to the structure host made of extended honeycomb layered systems [6]. Note that regarding *p*-elements magnetism, ordered magnetic moments were identified in hexaborides $AE$B$_6$ ($AE$ = Ca, Sr) [7] as well as in CdS doped with main group elements [8].

Focusing on carbon herein, one is presented with two possible situations regarding the onset of magnetization through:

i- Keeping an isolated atomic behavior -as Gd (4f)- characterized by two unpaired p electrons in an appropriate crystal-chemical host, and leading eventually to a polarization of the two p electrons and the expectation of M(C ) = 2 $\mu_B$. We assess this within an intraband spin polarization mechanism;

ii- Checking for the hypothesis of interband spin polarization mechanism by involving two carbon neighbors at a distance which allows electronic interactions between them. The subsequent bonding involves a spin pairing and a loss of the magnetization magnitude. The expectation is a magnitude of M(C ) close to 1 $\mu_B$.

Along with these two hypotheses, and based on DFT calculation of energies and magnetic configurations, the paper presents results and assessments of the two mechanisms in new carbon-based model multilayered systems.

**Brief presentation of the computational methodology**

Within the DFT, the optimization of the candidate structures (the atomic positions and the lattice parameters) is needed in the first place to identify the minimum energy configuration. The plane wave Vienna ab initio simulation package (VASP) package [9, 10] was used with its implementation of the projector augmented wave (PAW) method [10, 11]. The DFT exchange-correlation effects were accounted for with the generalized gradient approximation (GGA) scheme [12]. A conjugate-gradient algorithm [13] was used to relax the atom positions of the different compositions into the ground-state structure. Structural parameters were considered as fully minimized when forces on the atoms were less than 0.02 eV/Å and the stress components were below 0.003 eV/Å$^3$. A tetrahedron method [14] was applied for geometry relaxation and total energy calculations. The integrals within the reciprocal space (Brillouin-zone BZ) were approximated using a special **k**-point sampling [15]. The calculations were converged at an energy cut-off of 400 eV for all compounds. The **k**-mesh integration was carried out with increasing BZ precision over successive calculations for best convergence and relaxation to zero strains. Calculations were systematically carried out considering both non-spin-polarized (NSP) and spin-polarized (SP) –magnetic configurations. We also considered an electron localization EL mapping from real-space analysis of EL function (ELF) according to Becke and Edgecomb [16]. ELF is a normalized function with $0 \leq ELF \leq 1$, ranging from 0 for no localization to 1 for full localization; magnitudes ~1/2 correspond to free-electron like localization. In ELF maps the corresponding color scheme follows: blue areas for ELF=0; red color for ELF~1 and green color for ELF ~ ½ (cf. Figs. 2).

**Results and discussion.**

*Geometry optimization and energy-dependent results.*

$CC'_n (n= 6, 12, 18)$

Honeycomb carbon networks were identified in the lithium graphitic anode materials $LiC_6$ and $LiC_{12}$ [17] with the *P*6/*mmm* space group. The respective structures transposed to $CC'_6$ and $CC'_{12}$ are shown in Figs 1 a) and b). Full geometry relaxations were carried out assuming NSP and SP configurations. The results are given in Table 1) and b). In both compounds, the SP configuration has lower energy than the NSP and the cohesive energies obtained from subtracting the atomic C contributions are negative with the same trends of larger stability of SP. Comparing $CC'_6$ to $CC'_{12}$, the latter exhibits a larger cohesive energy letting propose that two C'6 layered compounds are more favorable. One can also note that the magnetization of M= 2 $\mu_B$ remains the same in the two compounds, due to the full polarization of carbon two p electrons. Regarding the crystal parameters, the onset of spin polarization causes a large increase of the hexagonal *c/a* ratio while few changes can be observed for the hexagonal *a* lattice constant as well as the C' positions which are mainly determined by the rigid C'6 network. In both stoichiometries, d(C'-C') changes little with ~1.4 Å and little affected by the spin state. Such trends and observations needed to be further confirmed with a structure comprising three C'6 layers with the formulation CC'''$_{18}$, with C at the origin and the layers at C' and C". The obtained structure within the same space group is shown in Fig. 1c and the geometry optimized parameters and energies explicated in Table 1c. Indeed the trend of stabilization through multilayered stoichiometry is confirmed with similar trends of lattice parameters and an increase of hexagonal *c/a* ratio always larger in SP than with the increase of the $E_{coh.}$ magnitude to -1.69 eV, i.e. versus $E_{coh.}$ ($CC'_{12}$) = -1.518 eV and $E_{coh.}$($CC'_6$) = -1.021 eV (SP values). Also, the SP configuration is more cohesive than NSP and the magnitude of the magnetic moment on C is 1.98 $\mu_B$, close to saturation moment of 2 $\mu_B$.

A most relevant result from the calculations is the integer ~2 $\mu_B$ magnitude of corner C in all structures. The C-C separation is large and amounts to the *a* lattice constant magnitude which is systematically larger than 4 Å, letting suggests an isolated character and a development of onsite magnetic moment through intraband spin polarization.

*$C_2C'_{12}$*

Support of the above hypothesis was needed with a structure accounting for two neighboring carbon atoms that would magnetically polarize through an inter-band mechanism.

Fig. 1 d) shows $C_2C'_{12}$ structure based on the *P6/mmm* space group and where C belongs to a two-fold Wyckoff position, i.e. at 1/3, 2/3, ½ and 2/3, 1/3, ½. After full geometry relaxation, the C-C distance is 2.46 Å. The results show SP as the ground state with, however, cohesive energy of -1.22 eV/at, smaller than the cohesive energy of $CC'_{12}$ of -1.518 eV. The remarkable result is now the lowering of the magnitude of magnetization which amounts to 1.97 $\mu_B$ for the two carbon atoms, i.e., 0.985 $\mu_B$/C atom. The long-range magnetic order is then ferromagnetic. Further antiferromagnetic calculations imposing two magnetic substructure, one labeled as SPIN UP and the other as SPIN-DOWN, i.e. C↑…C↓ as well as all the C' atoms. At self-consistent convergence, a raise of the energy was noted and the system becomes less cohesive.

One can then assume that the loss of magnetization magnitude is due to spin pairing through the bonding between the two carbon atoms, leaving one nonbonding spin polarizing along the *c* hexagonal direction. The mechanism of spin polarization is then suggested to be of interband nature.

*Magnetic charge density*

The argumentations assessments need to be supported by the magnetic charge densities on the different atomic constituents on one hand and with a projection of the electron localization on the other hand.

Regarding the former, the colored volumes around the C atoms (corner) in Figs. 1 a) to c) and the z-oriented envelope around the two C atoms of $C_2C'_{12}$ (Fig. 1d) corresponds to the charge density difference between majority spins (↑) and minority spins (↓). Clearly for the C' and C" atoms belonging to the C'6 layers have no magnetic charge density around them since the difference of population ↑-↓ is zero. While this is expected it is interesting to note the difference of shapes between the spherical like volume for the isolated C in Figs 1 a) to c) versus the elongated shape along with the c direction in Fig. 1d).

*Electron localization function ELF*

Figure 2 shows the different structures with the ELF slices crossing the basal planes containing the C magnetic atoms. The ruler at the lower part of the figure indicates the code: 0 (blue zones; zero localization); ½ (green zones, free electron-like) and 1 (red: full localization). Firstly in Fig. 2a we show the ELF of a C' layer in 2×2×1 projection where the red zones of strong localization allow depicting the C'6 (C"6) rings forming such layers. There are no blue zones pointing to no electron localization and the layer is similar to a graphitic one letting suggest a semi-conducting like behavior for the layer. Focusing on the basal planes, Figs. 2 b) to d) exhibit blue zones of no localization between the C sites thus supporting their isolated character C—C electronically and magnetically. Oppositely, Fig. 2e relevant to $C_2C'_{12}$, the two C atoms are in the basal plane at a distance of 2.46 Å and there is a significant green zone of free electron-like character between the two atoms. Then the ELF projections provide further illustration of the different magnetic behaviors observed, i.e. intraband versus interband magnetic polarizations.

*Analysis of the electronic density of states*

Considering non-spin-polarized configurations (NSP), Figure 3 shows the orbital projected DOS of C typically observed in the $CC'_n$ stoichiometries. There are two kinds of partial DOS: an intense sharp peak and a slightly broader one, both centered on the Fermi level (zero of energy along the *x*-axis). In the *P*6/*mmm* space group, the corresponding point group is $D_{4h}$ which leads to the observed split in 2 manifolds $A_{1u}$ and $E_{2u}$ for $p_z$ and $p_{x,y}$ respectively. The high DOS at $E_F$ is a signal of instability in such spin degenerate configuration.

From the subsequent SP calculations, the total DOS are plotted for majority spins (↑) and minority spins (↓) in Fig. 4. The difference between the two populations provides a magnetic moment. The energy position of the Fermi level is indicated and shown to be at a minimum of minority spins $n(E_F ↓)$ observed for all three $CC'_n$ stoichiometries, whence the integer 2 $\mu_B$/C atom and the half-metallic ferromagnet character.

Different features appear for the last panel corresponding to the $C_2C'_{12}$ SP DOS which show less energy shift difference between the two spin DOS and broader and less shaped DOS. Also one can notice the absence of a trough at $E_F$ for either $n(E_F)$ ↑ or ↓ with and a rather metallic-like character.

**Conclusions**.

We have shown that magnetically polarized carbon is increasingly stabilized upon considering hosting multilayer new $CC'_n$ (n = 6, 12, 18) stoichiometries within the hexagonal *P*6/*mmm* space group. It has been highlighted the occurrence for C of a particular mechanism for the onset of magnetization identified as being of an intra-band type featuring an integer magnetization of ~2 $\mu_B$ thanks to the large separation between neighboring atoms. Considering further a stoichiometry with 2 C closer neighbors in $C_2C'_{12}$, an interband spin polarization mechanism was identified and illustrated with an electron localization projection showing a free electron-like localization between the two atoms. The magnetization drops then to less than 1 $\mu_B$ due to the pairing of one spin to ensure for the interaction between the two atoms. Needless to say that the layered-like host structures proposed herein are unique in supporting magnetic polarizations (inter- and intra-); that is, with respect to $LaNi_5$-type or $AlB_2$-type, both in *P6/mmm* space group, but presenting less extended networks.

Further investigations are needed to clearly identify the magnitude of the Curie order temperature. Efforts for preparing such multilayer new stoichiometries are underway in collaboration with Europe.

**Acknowledgment**. We acknowledge the use of VESTA software for the structure and charge density sketches [18]. Computations were carried out on Xeon workstations of $C^2M^2S$ (Computer Center for Materials and Molecular Sciences) of the Lebanese German University.

Table 1. Calculated results for lattice parameters and distances given in Å and energies in eV. Space group $P6/mmm$ N° 191. In the three CC'$_n$ crystal structures, C is at (1$a$) (0,0,0) and in C$_2$C'$_{12}$, C at (2$c$) 1/3, 2/3, 0 (cf. Figs. 1). Atomic energy: $E_C$ = -7.11 eV.

a) CC'$_6$

| Magn. Config. | NSP | SP |
|---|---|---|
| $a$ | 4.285 | 4.272 |
| $c/a$ | 0.933 | 1.183 |
| d(C'-C') | 1.41-1.43 | 1.42-1.43 |
| C'(6$k$) $x$, 0, ½ | 0.335 | 0.333 |
| Tot. Energy | -56.23 | -56.91 |
| $E_{coh.}$ /at. | -0.932 | -1.021 |
| M ($\mu_B$) | - | 2.01 |

b) CC'$_{12}$

| Magn. Config. | NSP | SP |
|---|---|---|
| $a$ | 4.278 | 4.265 |
| $c/a$ | 1.820 | 2.060 |
| d(C'-C') | 1.42-1.43 | 1.42-1.43 |
| C'(12$n$) $x$, 0, $z$ | 0.334/0.280 | 0.333/0.297 |
| Tot. Energy | -111.46 | -112.16 |
| $E_{coh.}$ /at. | -1.464 | -1.518 |
| M ($\mu_B$) | - | 2.01 |

c) CC'''$_{18}$ (CC'$_{12}$C''$_6$)

| Magn. Config. | NSP | SP |
|---|---|---|
| $a$ | 4.275 | 4.265 |
| $c/a$ | 2.57 | 2.68 |
| d(C'-C')(C''-C'') | 1.41-1.42 | 1.41-1.42 |
| C'(12$n$) $x$, 0, $z$ | 0.334/0.183 | 0.333/0.196 |
| C''(6$k$) $x$, 0, ½ | 0.333 | 0.334 |
| Tot. Energy | -166.57 | -167.22 |
| $E_{coh.}$ /at. | -1.657 | -1.691 |
| M ($\mu_B$) | - | 1.980 |

d) $C_2C'_{12}$

| Magn. Config. | NSP | SP |
|---|---|---|
| $a$ | 4.220 | 4.230 |
| $c/a$ | 2.238 | 2.458 |
| $d(C'-C')$ | 1.41 | 1.41 |
| $C'(12n)$ $x, 0, z$ | 0.333/0.319 | 0.333/0.308 |
| Tot. Energy | -116.46 | -116.60 |
| $E_{coh.}$ /at. | -1.20 | -1.22 |
| M ($\mu_B$) | - | 1.97 (2 C atoms) |

# Figures

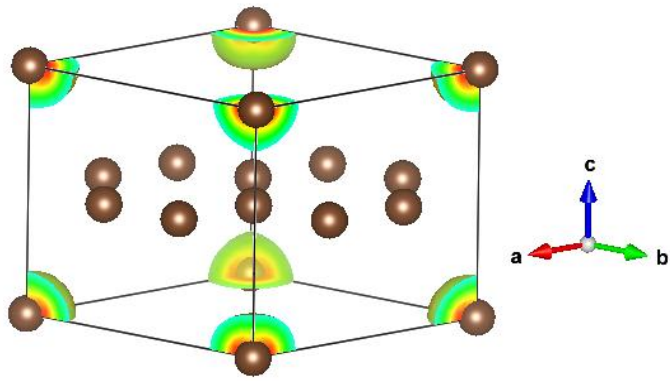

a) CC'$_6$

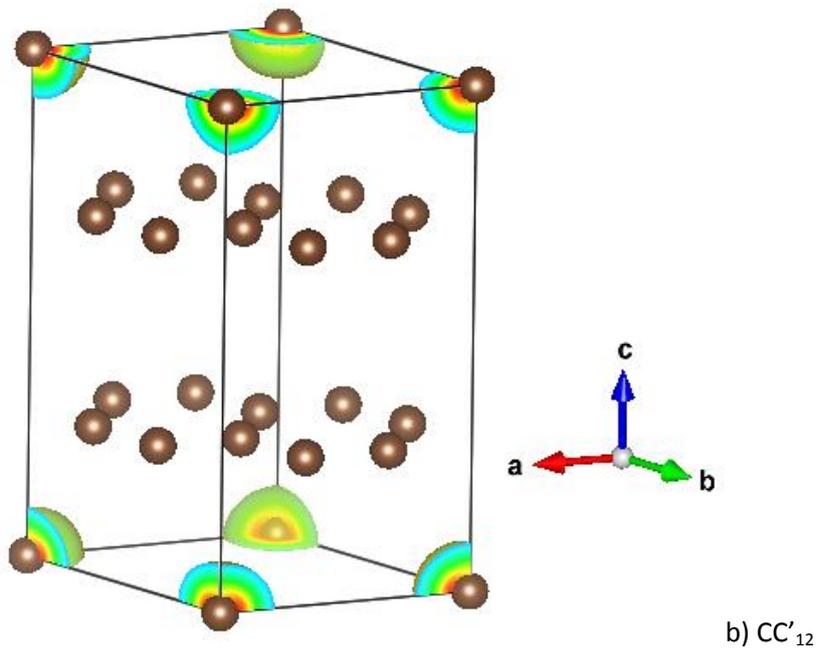

b) CC'$_{12}$

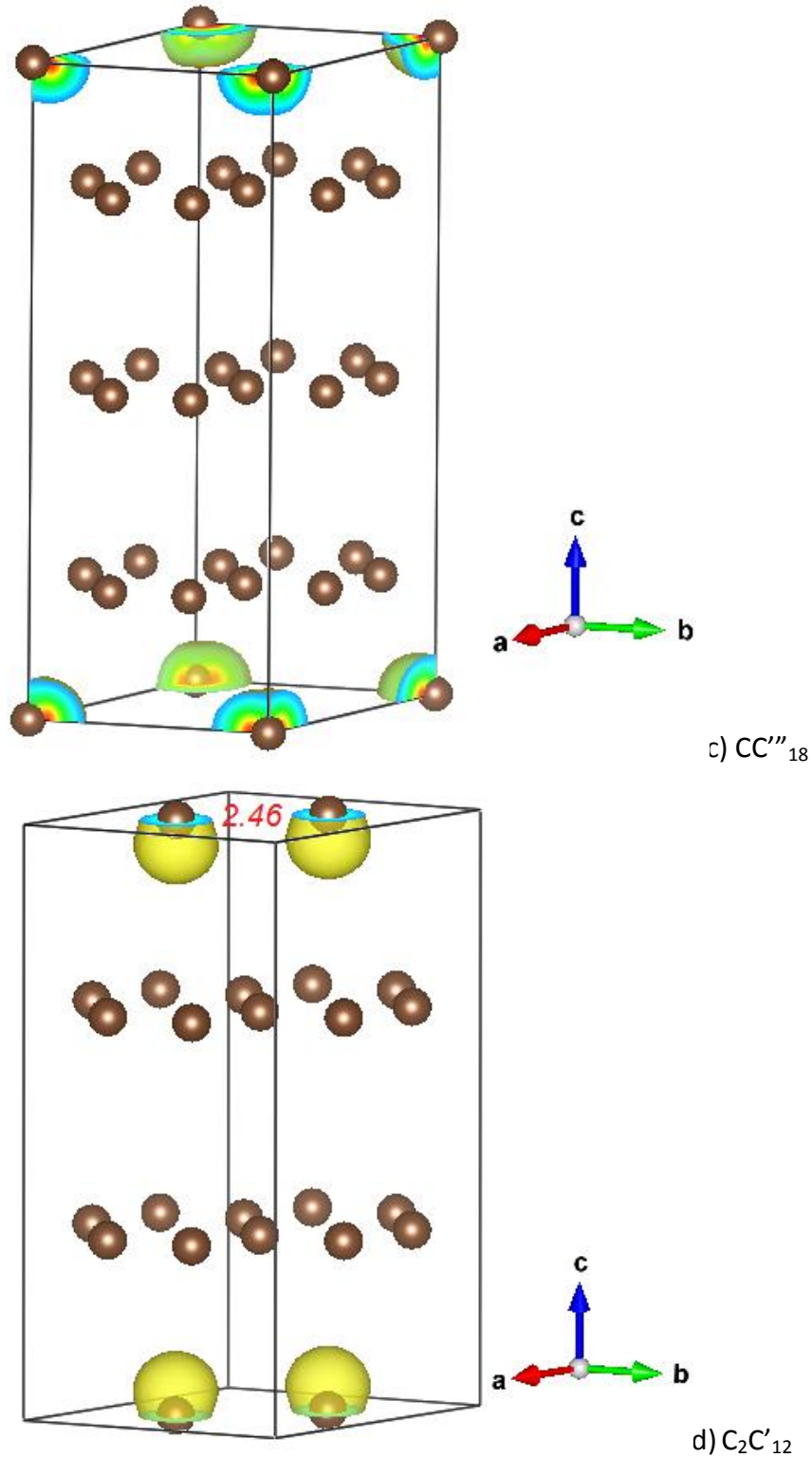

Figure 1. a) to c) Sketches of the crystal structures of mono, bi- and tri-layered chemical systems. Single carbon atoms at the corners are represented with magnetic charge density exhibited by the colored envelops, specifically showing that magnetic polarization occurs only on the corner carbon atoms C with M= 2 $\mu_B$. d) $C_2C'_{12}$ structure meant to show the prevailing interband spin polarization with a lowering to ~1 $\mu_B$ per carbon ith the presence of two C neighbors characterized by significant charge localization between them as shown in Fig. 2d).

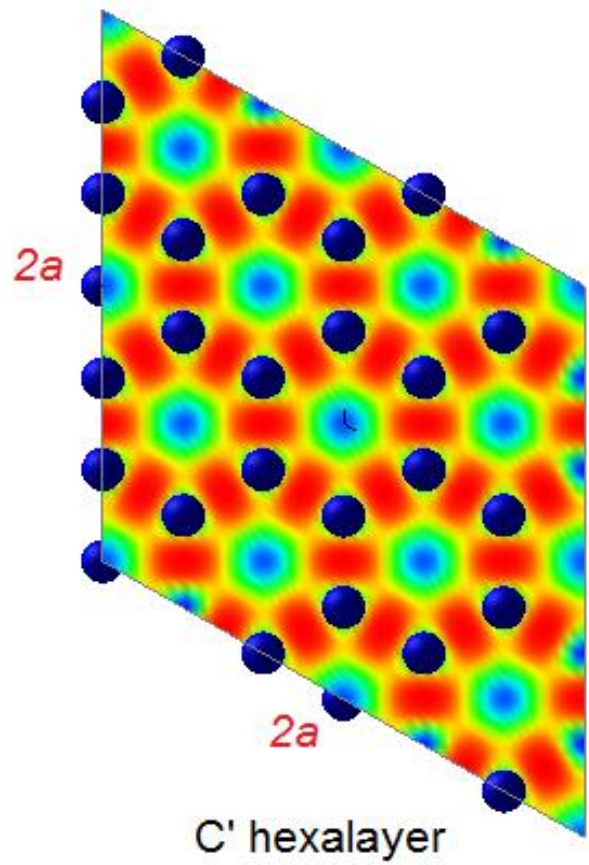

a) Carbon C′$_6$ layer over two adjacent cells.

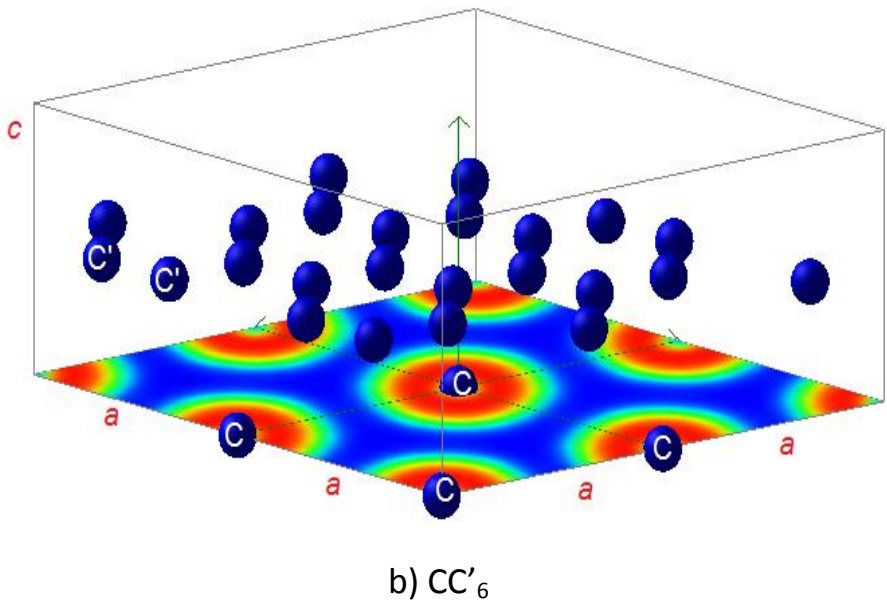

b) CC′$_6$

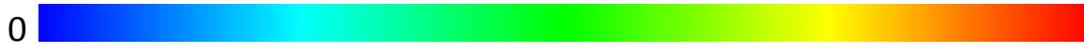

0

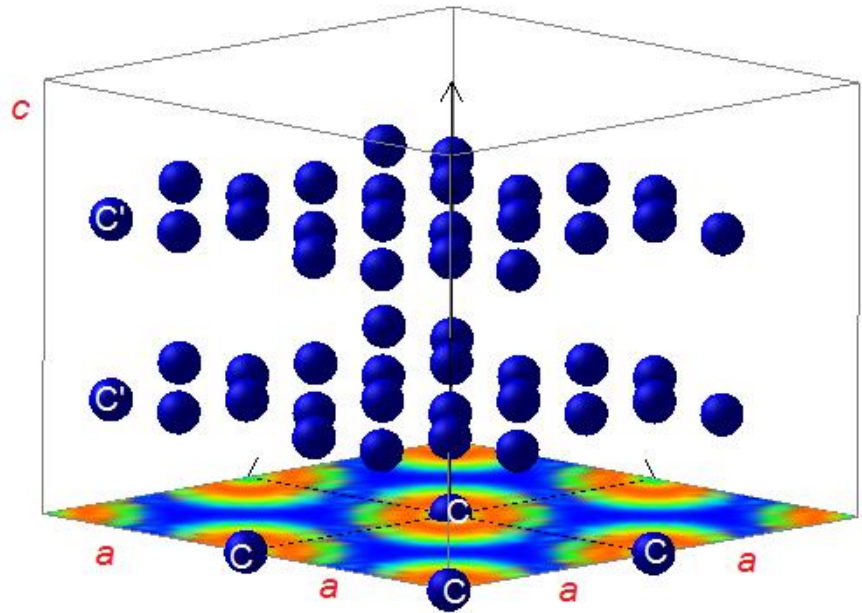

c) CC'$_{12}$

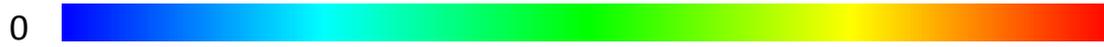

0

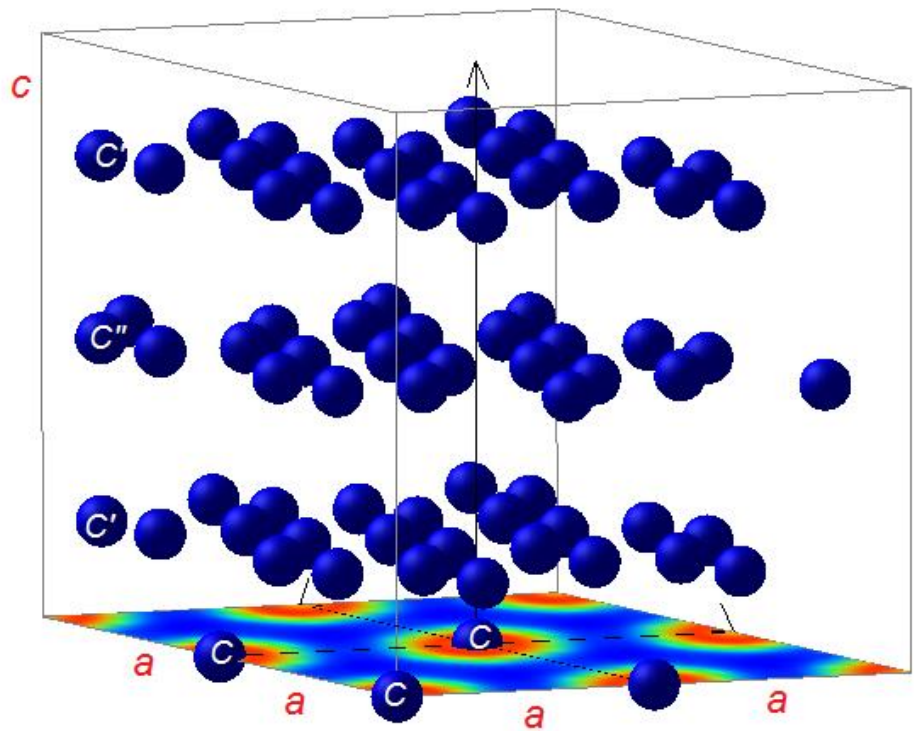

d) CC$_{18}$ (CC'$_{12}$ C''$_6$)

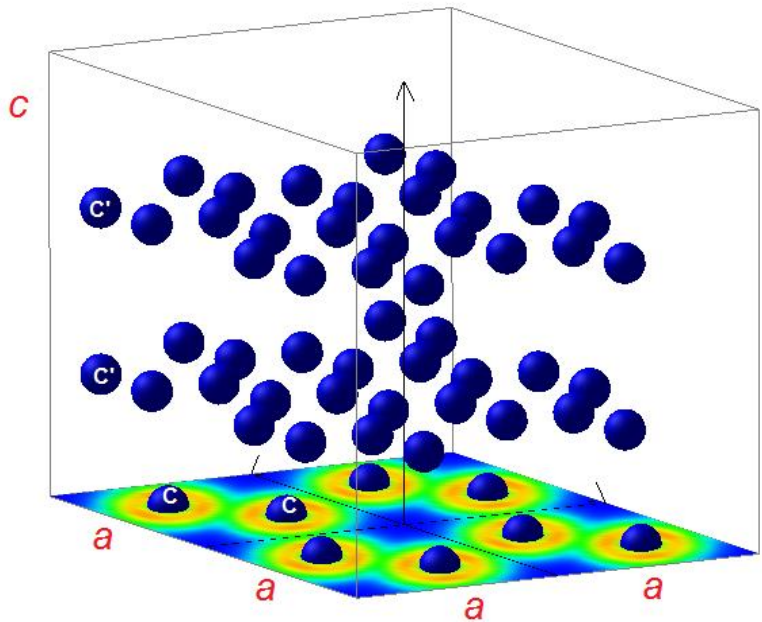

e) $C_2C'_{12}$

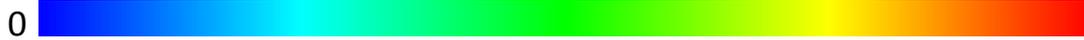
0

Figure 2. Electron localization ELF 2×2×1 projection. a) C'6 plane; b) to d) ELF slice planes crossing the basal plane containing C and e) the ELF crossing the plane containing two adjacent C atoms. The ruler shows the color code of ELF from 0 (blue: zero localization) to 1 (red: full localization). Notice the free electron-like behavior between the C atoms stressing the inter-band spin-polarization mechanism.

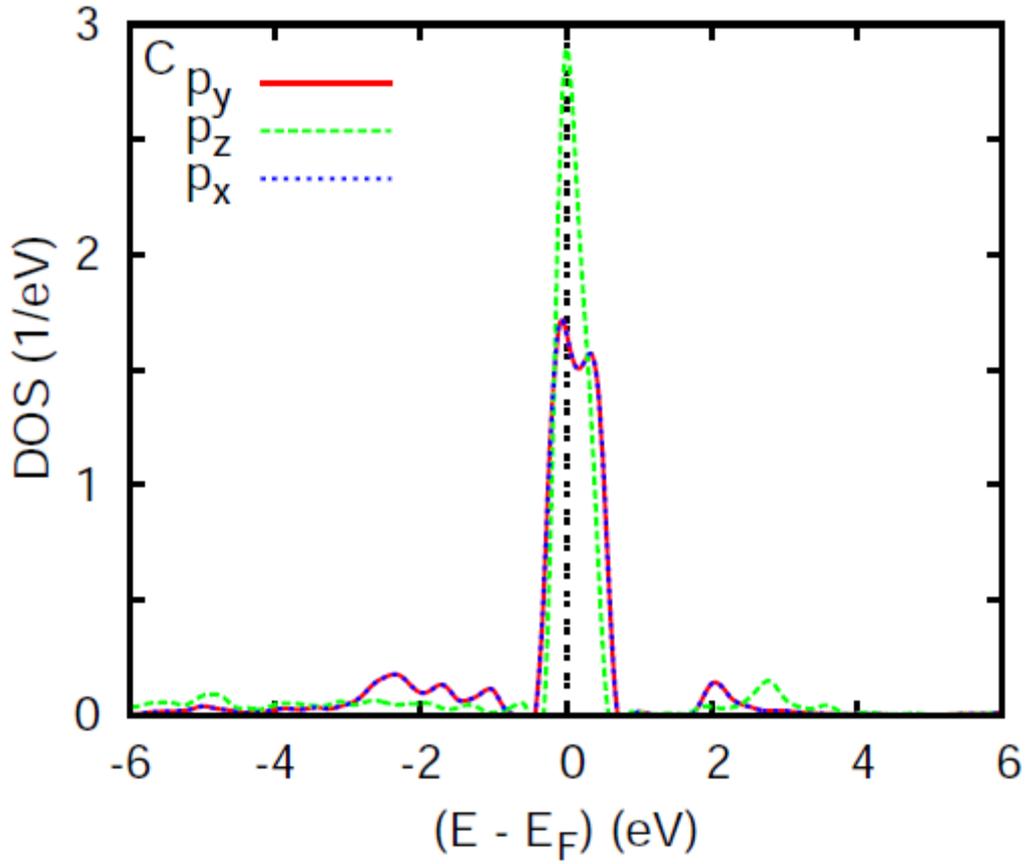

Figure 3. DOS of the three C p orbitals in NSP calculations of CC'$_n$. The splitting in the D$_{4h}$ point group of *P6/mmm* space group leads to the split in 2 manifolds: A$_{1u}$ and E$_{2u}$ for p$_z$ and p$_{x,y}$ resp. The high DOS at E$_F$ is a signal of instability in such spin degenerate configuration [1].

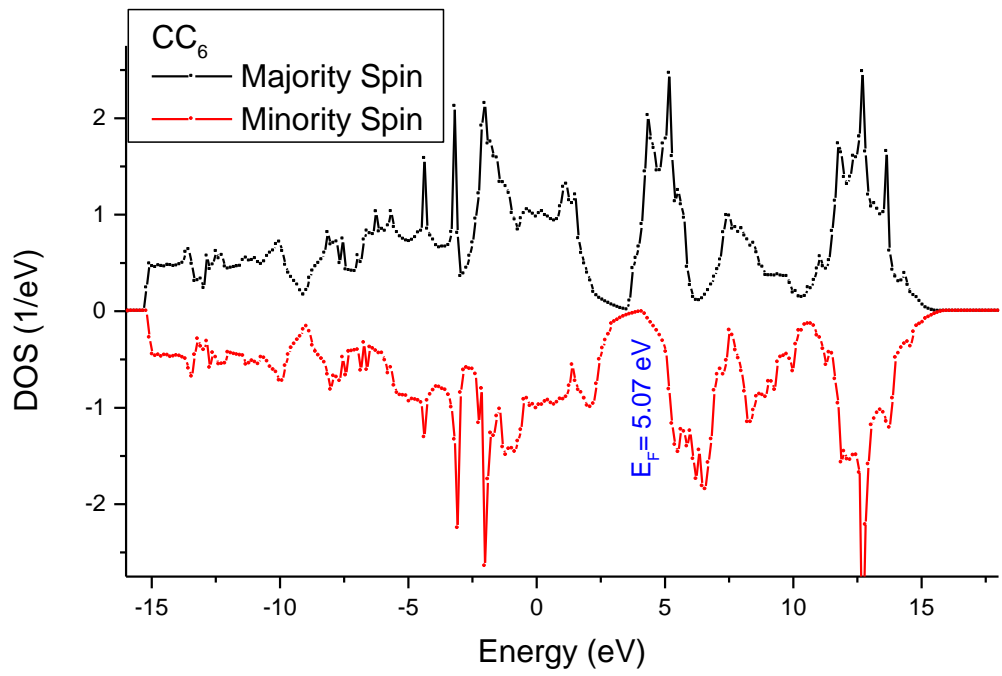

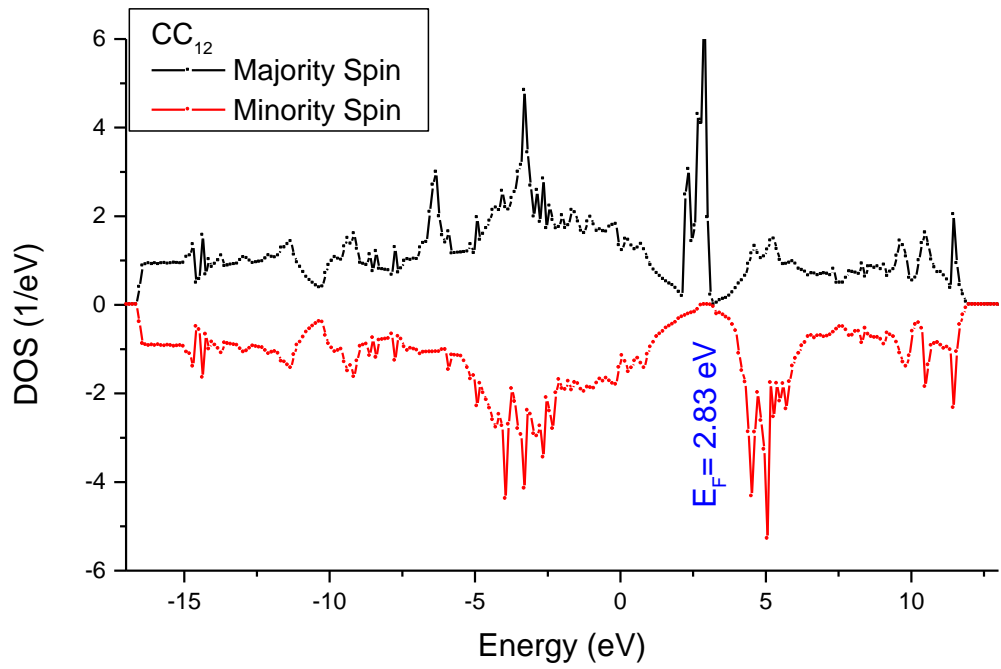

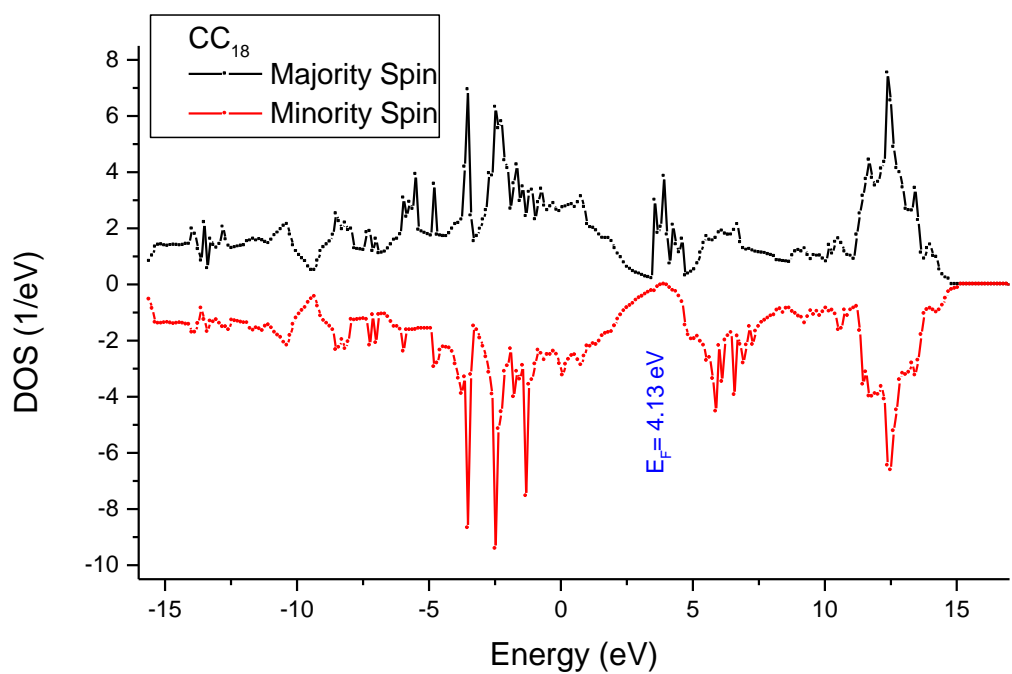

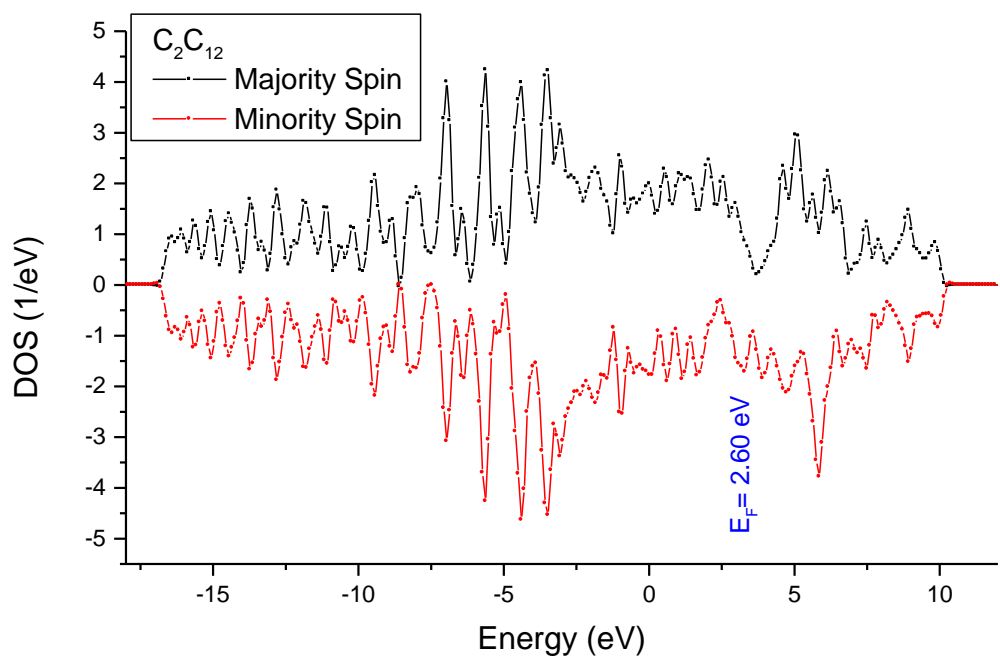

Figure 4: Spin projected total DOS of the $CC_n$ (n=6,12,18) and $C_2C_{12}$ stoichiometries.